\documentclass[%
 reprint,
nobibnotes,
bibnotes,
 amsmath,amssymb,
 aps,
]{revtex4-1}
\usepackage{MnSymbol}
\usepackage{natbib} 
\usepackage{xcolor}
\usepackage{soul}
\usepackage{hyperref}
\usepackage{wasysym} 
\usepackage{graphicx}
\usepackage{tikz}
\usetikzlibrary{calc}
\usepackage{dcolumn}
\usepackage{bm}

\tikzstyle{MyVertex}=[draw,circle,fill=black,inner sep=0,minimum size=3pt]
\tikzstyle{MyVacantVertex}=[draw,circle,fill=white,inner sep=0,minimum size=4pt]

\tikzstyle{MyDashedEdge}=[draw,dotted,line width=1pt]
\tikzstyle{MyFullEdge}=[line width=1pt]

\begin{document}

\preprint{Draft}


\title{Belief propagation on networks with cliques and chordless cycles}

\author{Peter Mann}
\email{pm78@st-andrews.ac.uk}
\author{Simon Dobson}
\affiliation{School of Computer Science, University of St Andrews, St Andrews, Fife KY16 9SX, United Kingdom }

\date{\today}

\begin{abstract}
    It is well known that tree-based theories can describe the properties of undirected clustered networks with extremely accurate results [S. Melnik, \textit{et al}. Phys. Rev. E 83, 036112 (2011)]. It is reasonable to suggest that a motif based theory would be superior to a tree one; since additional neighbour correlations are encapsulated in the motif structure. In this paper we examine bond percolation on random and real world networks using belief propagation in conjunction with edge-disjoint motif covers. We derive exact message passing expressions for cliques and chordless cycles of finite size. Our theoretical model gives good agreement with Monte Carlo simulation and offers a simple, yet substantial improvement on traditional message passing showing that this approach is suitable to study the properties of random and empirical networks.
\end{abstract}

\pacs{Valid PACS appear here}
\maketitle

\section{Introduction}
\label{sec:intro}

Belief propagation, also known as message passing, is an algorithm that is fundamental to many different disciplines including physics, computer science, epidemiology and statistics \cite{yedidia_freeman_weiss_2005,newman_2019,cantwell_newman_2019,kirkley_cantwell_newman_2021,mezard_parisi_zecchina_2002,PhysRevE.56.1357,michaelbeyond2022,bodnar2021weisfeiler}. The belief propagation algorithm, which relies on the Bethe-Peierls approximation \cite{bethe_1935,8262801,WellerEtAl_uai14}, is valid only for large and sparse treelike networks; becoming exact when the network \textit{is} a tree. This is because the removal of a vertex, $i$, and its edges from a tree isolates its neighbours from one another; the only path connecting them has vanished and so too do inter-neighbour correlations. Such a graph is called the \textit{cavity graph} of vertex $i$ and allows a message passing system of self-consistent equations to be written for the vertices of the network. Loops in networks introduce correlations between vertices that invalidates the Bethe-Peierls approximation as there is still a path between the neighbours in the cavity graph of a given vertex. Despite this drawback, belief propagation has a proven ability to analytically describe the structural properties of many real world networks, which are known to contain loops \cite{PhysRevLett.113.208702}. Belief propagation on clustered networks has been studied previously \cite{yedidia_freeman_weiss_2005,Gujrati_2001,PhysRevB.71.235119,PhysRevE.82.036101,PhysRevE.84.055101,PhysRevE.84.041144,PhysRevE.100.012314,PhysRevLett.113.208701,Coolen_2016,kirkley_cantwell_newman_2021,NIPS1997_0245952e, cantwell_newman_2019,PhysRevE.99.042309,michaelbeyond2022,bodnar2021weisfeiler,PhysRevE.73.065102} and an immense literature exists that tackles the problem in a variety of different manners. 

Recently, Cantwell, Newman and Kirkley solved the belief propagation model for networks with arbitrary loop structure \cite{kirkley_cantwell_newman_2021, cantwell_newman_2019}. They achieved this by considering messages from a collective local neighbourhood, of variable distance, about each vertex, rather than a partition into recognised motifs. In their framework, it is assumed that all inter-vertex correlations due to short range loops are encapsulated within the neighbourhood. For a given configuration of edges, the probability of the size of the component to which a vertex belongs is calculated, before being averaged over the probability of each edge configuration for the neighbourhood. The neighbourhood calculation is computationally expensive and so the authors introduced a Monte Carlo algorithm to sample the set of connected graphs. 

It is not unreasonable, however, to desire a \textit{model} of an empirical network to carry out further experimentation on. For instance, suppose that the empirical network represents a data set that is too small to apply some statistical or machine learning algorithm. If we had a model of the network, perhaps we could understand how to grow the network, adding new vertices and edges, whilst keeping the inherent topology and statistical properties fixed. 

\begin{figure}[ht!]
    \centering
    \includegraphics[width=0.45\textwidth]{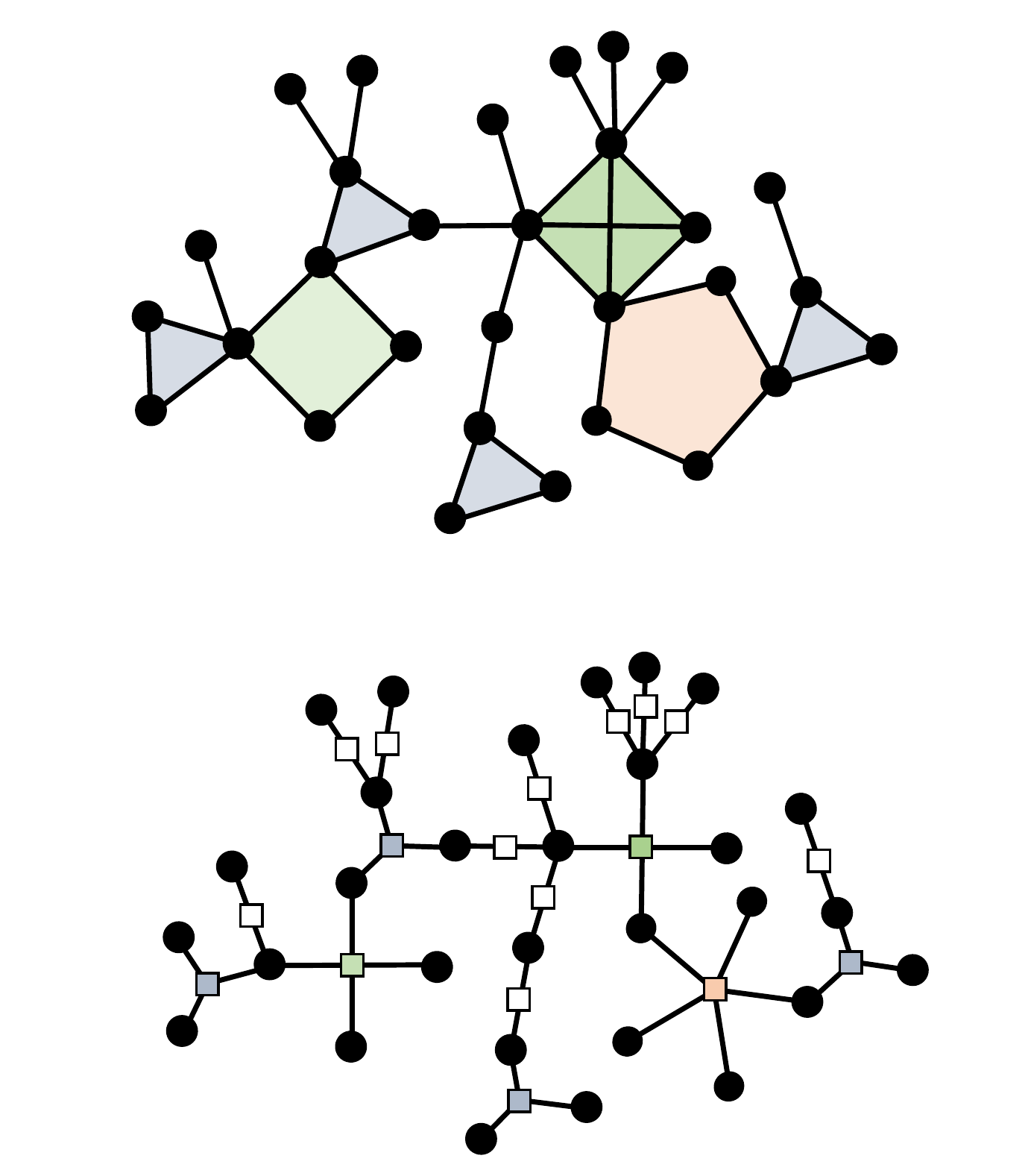}
    \caption{Top: A network is covered with edge-disjoint chordless cycles and cliques. Bottom: the factor graph of the network. Figure inspired by Figure 1 of \cite{PhysRevE.100.012314}. Such networks can have a high local density of loops that encapsulate the neighbour correlation, becoming increasingly sparse and treelike at long ranges.}
    \label{fig:factor_graph}
\end{figure}

One way to obtain a model is to cover the network in motifs; specifically, a set of \textit{edge-disjoint} motifs, such that a given edge belongs to only one motif. The simplest cover is to simply assume each edge is a 2-clique \cite{PhysRevLett.113.208702} - and this approximation certainly works well in some cases. However, given the large body of knowledge for random clustered networks \cite{PhysRevLett.103.058701,PhysRevE.80.020901,karrer_newman_2010,PhysRevE.101.062310,PhysRevE.105.044314,PhysRevE.104.024304,PhysRevE.103.012313,PhysRevE.103.012309,PhysRevE.84.041144,burgio_arenas_gomez_matamalas_2021,HASEGAWA2021125970} it is tempting to apply covers with larger and more complicated motifs in the hope to obtain more accurate models. For sparse random graphs, this technique works well and the success lies in the locally treelike nature of the \textit{factor graph} of the covered network. The factor graph is a bipartite graph that has two different sets of vertices. One set represents the vertices of the substrate network, whilst the other represents the motifs to which they belong. Edges connect vertices in the original graph to the vertices representing the motifs to which they belong, see Fig \ref{fig:factor_graph}. If the factor graph is locally treelike then all of the short range loops in the network are encapsulated within the motif cover. 

Unfortunately, naive motif covers of empirical networks often do not create representative models. Cantwell, Kirkley and Newman summarise that ``\textit{these techniques are not generally applicable to real world networks}'' \cite{kirkley_cantwell_newman_2021}. In fact, it is sometimes the case that tree-based theories give the closest match to the empirical network. This is despite a cover with larger motifs necessarily having a more treelike factor graph than a cover comprised of only 2-cliques. This indicates the presence of an additional driving force for a suitable motif cover beyond simply decreasing the number of loops in the factor graph. 

In this paper, we introduce exact expressions for belief propagation on random networks that are composed of chordless cycles and cliques of finite size. We then study message passing on empirical clique covered networks. We find that our model exhibits excellent agreement with Monte Carlo simulation beyond the results obtained from ordinary belief propagation \cite{PhysRevLett.113.208702} due to the arbitrarily large clique sizes we can analytically include in our model. We offer insight into why naive motif covers often fail to capture the properties of real world networks by considering the statistical bias introduced by the breaking of symmetry in certain covers. Finally, we relate our method to the generalised configuration model which is shown to be the ensemble average of the message passing model. 

This work will lead to significant advances in the application of the theory of clustered graphs to the study of empirical networks. We hope that additional studies will be conducted with dynamics other than bond percolation to further these results.


\section{Theoretical}
\label{sec:theoretical}

Let $G=(N,E)$ be a graph, and $\omega:E\to \{0,1\}$ be a \textit{edge configuration} on the graph that maps each edge to a value of either 0 or 1 with probabilities $1-\phi$ and $\phi$, respectively for $\phi\in [0,1]$. An edge is \textit{occupied} if $\omega(e)=1$ for $e\in E$, and \textit{unoccupied} if $\omega(e)=0$; as $\phi$ is increased a giant connected component of occupied edges emerges through a phase transition. The state space of the model is $\Omega=\{0,1\}^E$ such that $\omega\in \Omega$ are $E$-dimensional vectors. Let $\eta(\omega)=\{ e\in E : \omega(e)=1\}$ denote the set of occupied edges. For a given network $G$ the probability measure of an edge configuration is 
\begin{equation}
    \mu(\omega\mid G) =\frac{1}{Z} \prod_{e\in E}\phi^{\omega(e)}(1-\phi)^{\omega(e)},\label{eq:mu}
\end{equation}
where $Z$ is the partition function 
\begin{equation}
    Z=\sum_{\omega\in \Omega}\prod_{e\in E}\phi^{\omega(e)}(1-\phi)^{\omega(e)}.
\end{equation}
This summation is over all possible edge configurations of a given graph, weighted by their probabilities and therefore, is equal to one. When $\phi$ is small, the graph is composed of many small clusters of occupied vertices and there is no long range connectivity. As $\phi$ is increased a giant percolating cluster emerges which incorporates a finite fraction of the vertices $\mathcal O(N)$.  

We now derive the belief propagation formulation for motif covered networks. Similar expressions appear in \cite{cantwell_newman_2019} for vertex neighbourhood decompositions; and \cite{PhysRevE.84.041144} for the Ising model on motif covered networks. Let us select a vertex $i$ at random from the equilibrium of the percolation process and suppose that $i$ is the corner of a set of motifs $\bm \tau_i$. The probability $\pi_i(s)$ is the probability that $i$ belongs to a non-percolating component of size $s$. This can be written in terms of the probability $\pi_{i\leftarrow j}$ that a neighbour of $i$, vertex $j$, which is in motif $\tau$ together with $i$, leads to $s_j$ vertices if we were to follow all of its edges, other than those that point back to motif $\tau$
\begin{align}
    \pi_i(s) =&\  \sum_{\{s_j:j\in\partial_{\tau}(i), \ \tau\in \bm \tau_i\}}\left[\prod_{\tau\in \bm \tau_i}\prod_{j\in\partial_\tau(i)}\pi_{i\leftarrow j}(s_j)\right]\nonumber\\
    &\ \times\delta\left(s-1,\sum_{\tau\in \bm \tau_i}\sum_{j\in \partial_\tau(i)}s_j\right) ,
\end{align}
where $\tau$ is a motif belonging to $\bm \tau_i$,  $\partial_\tau(i)$ is the set of edges vertex $i$ has within motif $\tau$ and where $\delta(x,y)$ is the Kronecker delta. This expression averages over every combination of the total number of reachable vertices $s_j$ along each edge $j$ of each motif $\tau$ in the set of motifs $\bm \tau_i$ that $i$ belongs to. This summation is then filtered by the Kronecker delta to retain only those terms that collectively sum to $s-1$, to which we add 1 for vertex $i$ itself to yield the component of overall size $s$. We can generate this probability by defining
\begin{equation}
    G_i(z)=\sum\limits_{s=1}^\infty\pi_i(s)z^s\label{eq:G_i_main}
\end{equation} 
such that
\begin{align}
    G_i(z)=&\ \sum_{s=1}^\infty z^s  \sum_{\{s_j:j\in\partial_{\tau}(i), \ \tau\in \bm \tau_i\}}\left[\prod_{\tau\in \bm \tau_i}\prod_{j\in\partial_\tau(i)}\pi_{i\leftarrow j}(s_j)\right]\nonumber\\
    &\ \times\delta\left(s-1,\sum_{\tau\in \bm \tau_i}\sum_{j\in \partial_\tau(i)}s_j\right) .
\end{align}
With the following manipulation
\begin{equation}
    \sum^\infty_{s=1}z^s=z\sum^\infty_{s=1}z^{s-1} \stackrel{\delta}{=} z\sum^\infty_{s=1}z^{\sum\limits_\tau\sum\limits_{j\in\partial_\tau(i)}s_j}=z\prod_{\tau\in \bm \tau_i}\prod_{j\in\partial _\tau(i)}\sum^\infty_{s_j=0}z^{s_j},\label{eq:zlogic}
\end{equation}
we obtain 
\begin{equation}
    G_i(z) = z\prod _{\tau\in \bm \tau_i}\prod_{j\in \partial_\tau(i)}\sum^\infty_{s_j=0}\pi_{i\leftarrow j} (s_j)z^{s_j}.\label{eq:tempGi}
\end{equation}
The summation limits over $s_j$ are $0\rightarrow \infty$; since, the original summation is over $s\in [1,\infty]$ and so $s-1$ is $s_j\in[0,\infty]$. For each neighbour $j$ we can write a generating function for the probability that the number of vertices that can be reached, other than through the motif itself, is $s_j$ as
\begin{equation}
    H_{i\leftarrow j}(z)=\sum^\infty_{s_j=0}\pi_{i\leftarrow j}(s_j)z^{s_j}\label{eq:H}.
\end{equation}
 The probability for the total number of vertices that can be reached via motif $\tau$ is simply the product
 \begin{equation}
      H_{i\leftarrow \tau}(z) =  \prod_{j\in\partial_\tau(i)}H_{i\leftarrow j}(z).
 \end{equation}
Inserting these expressions in Eq \ref{eq:tempGi} we find
\begin{equation}
    G_i(z) = z\prod _{\tau\in \bm \tau_i} H_{i\leftarrow \tau}(z). \label{eq:G_i}
\end{equation}
This expression is a general result and holds for arbitrary motif topologies. If we had knowledge of each $H_{i\leftarrow \tau}(z)$ this generating function would yield the distribution of component sizes that vertex $i$ belongs to. From this, for instance, we can find the size of the percolating component $S$ as one minus the probability that all vertices $i\in N$ belong to finite components
\begin{equation}
    S = 1 - \frac{1}{N}\sum_i\prod _{\tau\in \bm \tau_i} H_{i\leftarrow \tau}(1).\label{eq:S}
\end{equation}
The average size of the finite components $\langle s_i\rangle$ is given by
\begin{equation}
    \langle s_i\rangle = \frac{\sum\limits_{s}s\pi_i(s)}{\sum\limits_{s}\pi_i(s)}=\frac{G_i'(1)}{G_i(1)},
\end{equation}
where we have used Eq \ref{eq:G_i_main}. Inserting Eq \ref{eq:G_i} we have 
\begin{align}
     \langle s_i\rangle =1+\sum_{\tau\in \bm \tau_i}\frac{H_{i\leftarrow \tau}'(1)}{H_{i\leftarrow \tau}(1)},\label{eq:small_component}
\end{align}
where we have used 
\begin{align}
    \frac{d}{dz}\left[\prod_\tau H_{i\leftarrow \tau}(z)\right] =& \sum_\tau\left[\left(\frac{d}{dz}H_{i\leftarrow \tau}(z)\right)\prod_{\nu\neq\tau}H_{i\leftarrow\nu}(z)\right]\nonumber\\
    =&\left(\prod_\tau H_{i\leftarrow \tau}(z)\right)\left(\sum_\tau\frac{H'_{i\leftarrow\tau}(z)}{H_{i\leftarrow\tau}(z)}\right)\label{eq:derivative1}.
\end{align}
To progress we require the derivative of $H_{i\leftarrow \tau}(z)$.
However, the functional form of $H_{i\leftarrow \tau}(z)$, crucial to enumerating $G_i(z)$, depends on the structure of the motif and the configuration of its edges among the occupied and unoccupied states. We will now examine $H_{i\leftarrow \tau}(z)$ for ordinary edges, chordless cycles of arbitrary length and cliques of arbitrary size.

\subsection{Calculating $H_{i\leftarrow \tau}$ for ordinary edges}
\label{subsec:2-clique}
Let us for a moment assume that the network contains only 2-cliques - ordinary edges; which was previously studied by Karrer \textit{et al} \cite{PhysRevLett.113.208702}.  In this case Eq \ref{eq:G_i} reduces to 
\begin{equation}
    G_i(z)=z\prod_j H_{i\leftarrow j}(z),
\end{equation}
where the product over $j$ accounts for each distinct 2-clique $i$ belongs to. Because there is only a single edge in a 2-clique, this index equivalently runs over the neighbour vertices $H_{i\leftarrow \tau}(z)=H_{i\leftarrow j}(z)$. If the edge is unoccupied with probability $1-\phi$, then $\pi_{i\leftarrow j}(s)$ is zero. In this case, the edge does not contribute to the component size of $i$ and so $s=0$. If the edge is occupied with probability $\phi$, then $\pi_{i\leftarrow j}(s\geq 1)$ is non-zero. Therefore, the summation in Eq \ref{eq:H} has the zero term $\pi_{i\leftarrow j}(0)=1-\phi$ extracted, and for $s \geq 1$ we have
\begin{align}
    \pi_{i\leftarrow j}(s) =&\phi \sum_{\{s_k:k\in \partial_\nu(j)\backslash i\}}\left[\prod_{k\in \partial(j)\backslash i} \pi_{j\leftarrow k}(s_k) \right]\nonumber\\
    &\ \times\delta\left(s-1,\sum_{k\in\partial(j)\backslash i}s_k\right),
\end{align}
where the notation $\partial(j)\backslash i$ denotes the set edges to which vertex $j$ belongs to excluding the one connecting it to $i$.
Substituting this expression into Eq \ref{eq:H} we have 
\begin{align}
    H_{i\leftarrow j}(z) =&1-\phi+  \phi \sum_{s=1}^\infty z^s \sum_{\{s_k:k\in \partial (j)\backslash i\}}\left[\prod_{k\in \partial(j)\backslash i} \pi_{j\leftarrow k}(s_k) \right]\nonumber\\
    &\times\delta\left(s-1,\sum_{k\in\partial(j)\backslash i}s_k\right) .
\end{align}
Application of the Kronecker delta yields 
\begin{equation}
    H_{i\leftarrow j}(z) =(1 - \phi) +  \phi z \prod_{k\in \partial (j)\backslash i}\sum^\infty_{s_k=0}  \pi_{j\leftarrow k}(s_k)z^{s_k}.
\end{equation}
Noticing that the final summation is the generating function of $\pi_{j\leftarrow k}$ (Eq \ref{eq:H}) we have a self-consistent expression for $H_{i\leftarrow j}(z)$
\begin{equation}
     H_{i\leftarrow j}(z) =(1 - \phi) +  \phi z \prod_{k\in \partial (j)\backslash i}H_{j\leftarrow k}(z).\label{eq:htreenormal}
\end{equation}
This expression can be solved for all edges by fixed point iteration from a suitable starting point and substituted into Eq \ref{eq:G_i} to find $G_i(z)$. The distribution for $\pi_i(s)$ can then be found by differentiating the series. 

To evaluate the finite components, we require the derivative of $H_{i\leftarrow j}(z)$ with respect to $z$. Following \cite{PhysRevLett.113.208702} we find
\begin{equation}
    H'_{i\leftarrow j}(1) = \phi\left[1 + \sum_{k\in \partial (j)\backslash i}\frac{H'_{j\leftarrow k}(1)}{H_{j\leftarrow k}(1)}\right]\prod_{k\in \partial (j)\backslash i}H_{j\leftarrow k}(1).\label{eq:Htreederiv}
\end{equation}
To find the average size of the finite components we solve Eqs  \ref{eq:htreenormal} and \ref{eq:Htreederiv} and insert them into Eq \ref{eq:small_component}.
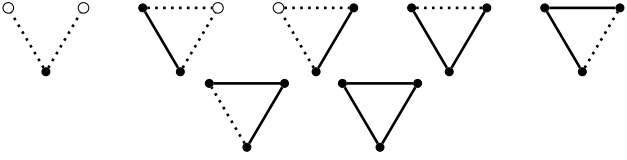
\begin{figure}
    \centering
     \begin{tikzpicture}[baseline=-0.25ex,scale=1]
    \node[MyVertex] (middle) at ( 0,   -0.2) {};
    \node[MyVacantVertex] (left)   at (-0.5, 0.65) {};
    \node[MyVacantVertex] (right)  at ( 0.5, 0.65) {};
    \draw[MyDashedEdge,black]
          (middle) -- (left);
    \draw[MyDashedEdge,black]
          (middle) -- (right);
\end{tikzpicture}\qquad 
   \begin{tikzpicture}[baseline=-0.25ex,scale=1]
    \node[MyVertex] (middle) at ( 0,   -0.2) {};
    \node[MyVertex] (left)   at (-0.5, 0.65) {};
    \node[MyVacantVertex] (right)  at ( 0.5, 0.65) {};
    \draw[MyFullEdge,black]
          (middle) -- (left);
    \draw[MyDashedEdge,black]
          (middle) -- (right);
    \draw[MyDashedEdge,black]
          (left) -- (right);
\end{tikzpicture}\qquad 
 \begin{tikzpicture}[baseline=-0.25ex,scale=1]
    \node[MyVertex] (middle) at ( 0,   -0.2) {};
    \node[MyVacantVertex] (left)   at (-0.5, 0.65) {};
    \node[MyVertex] (right)  at ( 0.5, 0.65) {};
    \draw[MyDashedEdge,black]
          (middle) -- (left);
    \draw[MyFullEdge,black]
          (middle) -- (right);
    \draw[MyDashedEdge,black]
          (left) -- (right);
\end{tikzpicture}\qquad
    \begin{tikzpicture}[baseline=-0.25ex,scale=1]
    \node[MyVertex] (middle) at ( 0,   -0.2) {};
    \node[MyVertex] (left)   at (-0.5, 0.65) {};
    \node[MyVertex] (right)  at ( 0.5, 0.65) {};
    \draw[MyFullEdge,black]
          (middle) -- (left);
    \draw[MyFullEdge,black]
          (middle) -- (right);
    \draw[MyDashedEdge,black]
          (left) -- (right);
\end{tikzpicture}\qquad \begin{tikzpicture}[baseline=-0.25ex,scale=1]
    \node[MyVertex] (middle) at ( 0,   -0.2) {};
    \node[MyVertex] (left)   at (-0.5, 0.65) {};
    \node[MyVertex] (right)  at ( 0.5, 0.65) {};
    \draw[MyFullEdge,black]
          (middle) -- (left);
    \draw[MyDashedEdge,black]
          (middle) -- (right);
    \draw[MyFullEdge,black]
          (left) -- (right);
\end{tikzpicture}\qquad 
\begin{tikzpicture}[baseline=-0.25ex,scale=1]
    \node[MyVertex] (middle) at ( 0,   -0.2) {};
    \node[MyVertex] (left)   at (-0.5, 0.65) {};
    \node[MyVertex] (right)  at ( 0.5, 0.65) {};
    \draw[MyDashedEdge,black]
          (middle) -- (left);
    \draw[MyFullEdge,black]
          (middle) -- (right);
    \draw[MyFullEdge,black]
          (left) -- (right);
\end{tikzpicture}\qquad 
\begin{tikzpicture}[baseline=-0.25ex,scale=1]
    \node[MyVertex] (middle) at ( 0,   -0.2) {};
    \node[MyVertex] (left)   at (-0.5, 0.65) {};
    \node[MyVertex] (right)  at ( 0.5, 0.65) {};
    \draw[MyFullEdge,black]
          (middle) -- (left);
    \draw[MyFullEdge,black]
          (middle) -- (right);
    \draw[MyFullEdge,black]
          (left) -- (right);
\end{tikzpicture}
    \caption{The seven edge configurations $\mathcal C$ that a focal vertex $i$ (bottom) can encounter when it belongs to a triangle with vertices $ j $ (top left vertex) and $ k $ (top right vertex). Solid lines represent occupied edges whilst dashed lines are unoccupied. Solid vertices belong to the same component as $i$, whilst unfilled vertices are not connected to $i$. When both of $i$'s edges are unoccupied (top left triangle) the state of the edge between the neighbours is inconsequential to the percolation properties of $i$. }
    \label{fig:triangles}
\end{figure}
\subsection{Calculating $H_{i\leftarrow \tau}$ for 3-cliques}
\label{subsec:3-clique}
Let us now examine the case where vertices belong to triangles as well as ordinary edges. Each triangle $\tau$ that vertex $i$ belongs to connects it to vertices $ j $ and $ k $; and, our aim is to calculate the generating function $H_{i\leftarrow \tau}(z)$ for the probability $\pi_{i\leftarrow \tau}(s_\tau)$ that the number of vertices that are reachable from vertex $i$ due to its membership in triangle $\tau$ is $s_\tau$. However, unlike ordinary edges, which contain just 1 edge, a triangle has 3 edges. Each of these edges may be occupied $\omega(e)=1$, or unoccupied $\omega(e)=0$ and there are 7 distinct edge configurations $\mathcal C$ for a given triangle that bear impact on the connectivity of vertex $i$, see Fig \ref{fig:triangles}. 

The probability that the triangle connects $i$ to $s_\tau$ vertices is dependent on the edge configuration that the triangle is in. Therefore, in order to compute $\pi_{i\leftarrow \tau}(s_\tau)$, we must calculate conditional probabilities $\pi_{i\leftarrow \tau}(s_\tau\mid \mathcal C)$ of leading to $s_\tau$ vertices for a given edge configuration $\mathcal C$. We find
\begin{widetext}
\begin{subequations}
\begin{align}
\pi_{i\leftarrow \tau}\left(s_\tau\ \big|\ 
    \begin{tikzpicture}[baseline=-0.25ex,scale=0.5]
    \node[MyVertex] (middle) at ( 0,   -0.2) {};
    \node[MyVacantVertex] (left)   at (-0.5, 0.65) {};
    \node[MyVacantVertex] (right)  at ( 0.5, 0.65) {};
    \draw[MyDashedEdge,black]
          (middle) -- (left);
    \draw[MyDashedEdge,black]
          (middle) -- (right);
\end{tikzpicture}
\right) =&\ \delta(s_\tau,0),\label{eq:subdelta}\\
\pi_{i\leftarrow \tau}\left(s_\tau\ \big|\ 
    \begin{tikzpicture}[baseline=-0.25ex,scale=0.5]
    \node[MyVertex] (middle) at ( 0,   -0.2) {};
    \node[MyVertex] (left)   at (-0.5, 0.65) {};
    \node[MyVacantVertex] (right)  at ( 0.5, 0.65) {};
    \draw[MyFullEdge,black]
          (middle) -- (left);
    \draw[MyDashedEdge,black]
          (middle) -- (right);
    \draw[MyDashedEdge,black]
          (left) -- (right);
\end{tikzpicture}
\right) =&\ \sum_{\{s_\ell:\ell\in \partial_\nu( j ),\ \nu \in \bm \nu_{ j }\backslash \tau\}} \left[\prod_{\nu\in \bm \nu_{ j }\backslash \tau} \prod_{\ell\in\partial_\nu( j )}\pi_{ j \leftarrow \ell}(s_\ell)\right]\delta\left(s_\tau-1,\sum_{\nu\in \bm \nu_{ j }\backslash \tau} \sum_{\ell\in\partial_\nu( j )}s_\ell\right)\label{eq:subtj},\\
\pi_{i\leftarrow \tau}\left(s_\tau\ \big|\ 
    \begin{tikzpicture}[baseline=-0.25ex,scale=0.5]
    \node[MyVertex] (middle) at ( 0,   -0.2) {};
    \node[MyVacantVertex] (left)   at (-0.5, 0.65) {};
    \node[MyVertex] (right)  at ( 0.5, 0.65) {};
    \draw[MyDashedEdge,black]
          (middle) -- (left);
    \draw[MyFullEdge,black]
          (middle) -- (right);
    \draw[MyDashedEdge,black]
          (left) -- (right);
\end{tikzpicture}
\right) =& \sum_{\{s_r:r\in \partial_\varphi( k ),\ \varphi \in \bm \varphi_{ k }\backslash \tau\}} \left[\prod_{\varphi\in \bm \varphi_{ k }\backslash \tau} \prod_{r\in\partial_\varphi( k )}\pi_{ k \leftarrow r}(s_r)\right]\delta\left(s_\tau-1,\sum_{\varphi\in \bm \varphi_{ k }\backslash \tau} \sum_{r\in\partial_\varphi( k )}s_r\right),\label{eq:subtk}
\end{align}
\end{subequations}
where the notation $\bm\nu_{ j }\backslash \tau$ in Eq \ref{eq:subtj} denotes the set of motifs that vertex $ j $ belongs to excluding $\tau$; and similarly for $\bm\varphi_{ k }\backslash \tau$ in Eq \ref{eq:subtk}. When the focal vertex is connected to \textit{both} $ j $ and $ k $ in $\tau$ via occupied edges, we must average over both sets of neighbours, see Fig \ref{fig:triangle_cavities}. 
\begin{figure}[ht!]
    \centering
    \includegraphics[width=0.425\textwidth]{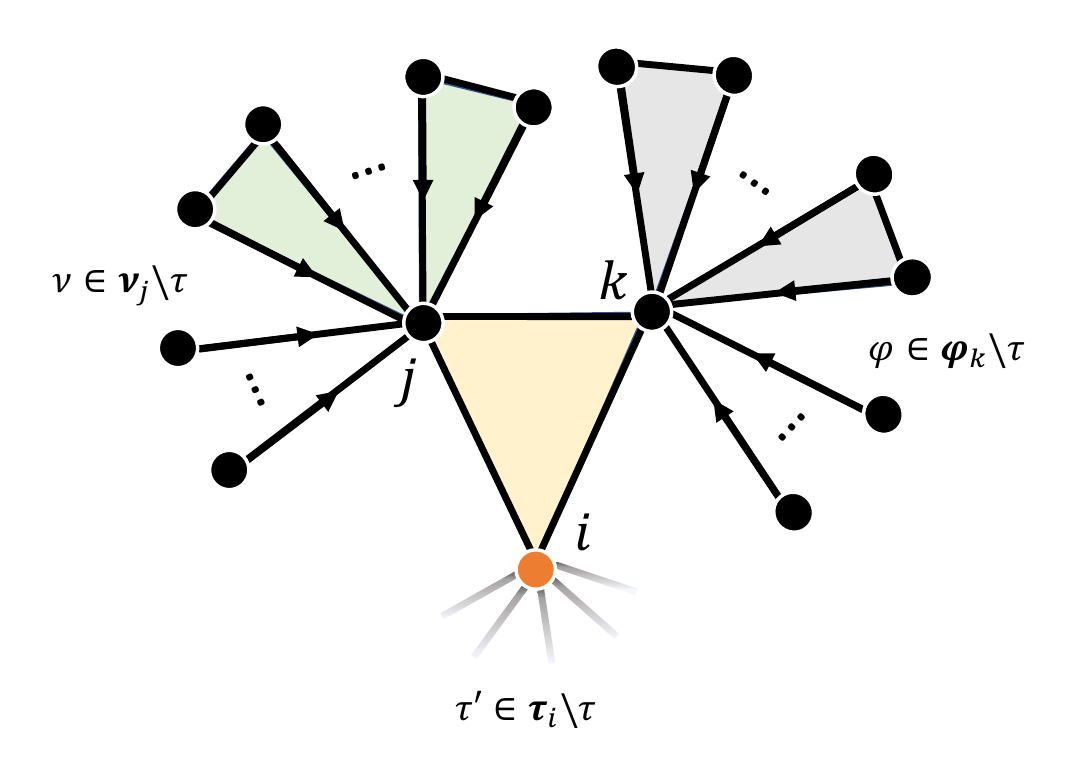}
    \caption{The probability that triangle $\tau$ connects vertex $i$ to $s_\tau$ vertices, when both $ j $ and $ k $ belong to the same component as $i$, is equal to the probability that the number of neighbouring vertices of $ j $ and $ k $, apart from those in $\tau$ itself, collectively connect to $s_\tau-2$ vertices, see Eq \ref{eq:double}. }
    \label{fig:triangle_cavities}
\end{figure}

\begin{align}
\pi_{i\leftarrow \tau}\left(s_\tau\ \big|\ 
    \begin{tikzpicture}[baseline=-0.25ex,scale=0.5]
    \node[MyVertex] (middle) at ( 0,   -0.2) {};
    \node[MyVertex] (left)   at (-0.5, 0.65) {};
    \node[MyVertex] (right)  at ( 0.5, 0.65) {};
    \draw[MyFullEdge,black]
          (middle) -- (left);
    \draw[MyFullEdge,black]
          (middle) -- (right);
    \draw[MyDashedEdge,black]
          (left) -- (right);
\end{tikzpicture},    \begin{tikzpicture}[baseline=-0.25ex,scale=0.5]
    \node[MyVertex] (middle) at ( 0,   -0.2) {};
    \node[MyVertex] (left)   at (-0.5, 0.65) {};
    \node[MyVertex] (right)  at ( 0.5, 0.65) {};
    \draw[MyFullEdge,black]
          (middle) -- (left);
    \draw[MyDashedEdge,black]
          (middle) -- (right);
    \draw[MyFullEdge,black]
          (left) -- (right);
\end{tikzpicture},
\begin{tikzpicture}[baseline=-0.25ex,scale=0.5]
    \node[MyVertex] (middle) at ( 0,   -0.2) {};
    \node[MyVertex] (left)   at (-0.5, 0.65) {};
    \node[MyVertex] (right)  at ( 0.5, 0.65) {};
    \draw[MyDashedEdge,black]
          (middle) -- (left);
    \draw[MyFullEdge,black]
          (middle) -- (right);
    \draw[MyFullEdge,black]
          (left) -- (right);
\end{tikzpicture},
\begin{tikzpicture}[baseline=-0.25ex,scale=0.5]
    \node[MyVertex] (middle) at ( 0,   -0.2) {};
    \node[MyVertex] (left)   at (-0.5, 0.65) {};
    \node[MyVertex] (right)  at ( 0.5, 0.65) {};
    \draw[MyFullEdge,black]
          (middle) -- (left);
    \draw[MyFullEdge,black]
          (middle) -- (right);
    \draw[MyFullEdge,black]
          (left) -- (right);
\end{tikzpicture}
\right) =&\  \sum_{\{s_\ell:\ell\in \partial_\nu( j ),\ \nu \in \bm \nu_{ j }\backslash \tau\}} \sum_{\{s_r:r\in \partial_\varphi( k ),\ \varphi \in \bm \varphi_{ k }\backslash \tau\}}\nonumber\\
&\times\left[\prod_{\nu\in \bm \nu_{ j }\backslash \tau} \prod_{\ell\in\partial_\nu( j )}\pi_{ j \leftarrow \ell}(s_\ell)\prod_{\varphi\in \bm \varphi_{ k }\backslash \tau} \prod_{r\in\partial_\varphi( k )}\pi_{ k \leftarrow r}(s_r)\right]\nonumber\\
&\times \delta\left(s_\tau-2,\sum_{\nu\in \bm \nu_{ j }\backslash \tau} \sum_{\ell\in\partial_\nu( j )}s_\ell+
\sum_{\varphi\in \bm \varphi_{ k }\backslash \tau} \sum_{r\in\partial_\varphi( k )}s_r
\right)\label{eq:double}.
\end{align}
The total probability that triangle $\tau$ leads to $s_\tau$ vertices is found by summing each conditional probability with the probability of each edge configuration \cite{PhysRevLett.103.058701,karrer_newman_2010,PhysRevE.104.024304}
\begin{align}
    \pi_{i\leftarrow \tau}(s_\tau) =&\ (1-\phi)^2\pi_{i\leftarrow  j }\left(0\ \big|\ 
    \begin{tikzpicture}[baseline=-0.25ex,scale=0.5]
    \node[MyVertex] (middle) at ( 0,   -0.2) {};
    \node[MyVacantVertex] (left)   at (-0.5, 0.65) {};
    \node[MyVacantVertex] (right)  at ( 0.5, 0.65) {};
    \draw[MyDashedEdge,black]
          (middle) -- (left);
    \draw[MyDashedEdge,black]
          (middle) -- (right);
\end{tikzpicture}
\right)+
\phi(1-\phi)^2\left[
    \pi_{i\leftarrow  j }\left(s_\tau\ \big|\ 
    \begin{tikzpicture}[baseline=-0.25ex,scale=0.5]
    \node[MyVertex] (middle) at ( 0,   -0.2) {};
    \node[MyVertex] (left)   at (-0.5, 0.65) {};
    \node[MyVacantVertex] (right)  at ( 0.5, 0.65) {};
    \draw[MyFullEdge,black]
          (middle) -- (left);
    \draw[MyDashedEdge,black]
          (middle) -- (right);
    \draw[MyDashedEdge,black]
          (left) -- (right);
\end{tikzpicture}
\right)+
\pi_{i\leftarrow  k }\left(s_\tau\ \big|\ 
    \begin{tikzpicture}[baseline=-0.25ex,scale=0.5]
    \node[MyVertex] (middle) at ( 0,   -0.2) {};
    \node[MyVacantVertex] (left)   at (-0.5, 0.65) {};
    \node[MyVertex] (right)  at ( 0.5, 0.65) {};
    \draw[MyDashedEdge,black]
          (middle) -- (left);
    \draw[MyFullEdge,black]
          (middle) -- (right);
    \draw[MyDashedEdge,black]
          (left) -- (right);
\end{tikzpicture}
\right)\right]\nonumber\\
&\ +
\left[3\phi^2(1-\phi)+\phi^3\right]
\pi_{i\leftarrow \tau}\left(s_\tau\ \big|\ 
    \begin{tikzpicture}[baseline=-0.25ex,scale=0.5]
    \node[MyVertex] (middle) at ( 0,   -0.2) {};
    \node[MyVertex] (left)   at (-0.5, 0.65) {};
    \node[MyVertex] (right)  at ( 0.5, 0.65) {};
    \draw[MyFullEdge,black]
          (middle) -- (left);
    \draw[MyFullEdge,black]
          (middle) -- (right);
    \draw[MyDashedEdge,black]
          (left) -- (right);
\end{tikzpicture},    \begin{tikzpicture}[baseline=-0.25ex,scale=0.5]
    \node[MyVertex] (middle) at ( 0,   -0.2) {};
    \node[MyVertex] (left)   at (-0.5, 0.65) {};
    \node[MyVertex] (right)  at ( 0.5, 0.65) {};
    \draw[MyFullEdge,black]
          (middle) -- (left);
    \draw[MyDashedEdge,black]
          (middle) -- (right);
    \draw[MyFullEdge,black]
          (left) -- (right);
\end{tikzpicture},
\begin{tikzpicture}[baseline=-0.25ex,scale=0.5]
    \node[MyVertex] (middle) at ( 0,   -0.2) {};
    \node[MyVertex] (left)   at (-0.5, 0.65) {};
    \node[MyVertex] (right)  at ( 0.5, 0.65) {};
    \draw[MyDashedEdge,black]
          (middle) -- (left);
    \draw[MyFullEdge,black]
          (middle) -- (right);
    \draw[MyFullEdge,black]
          (left) -- (right);
\end{tikzpicture},
\begin{tikzpicture}[baseline=-0.25ex,scale=0.5]
    \node[MyVertex] (middle) at ( 0,   -0.2) {};
    \node[MyVertex] (left)   at (-0.5, 0.65) {};
    \node[MyVertex] (right)  at ( 0.5, 0.65) {};
    \draw[MyFullEdge,black]
          (middle) -- (left);
    \draw[MyFullEdge,black]
          (middle) -- (right);
    \draw[MyFullEdge,black]
          (left) -- (right);
\end{tikzpicture}
\right).
\end{align}
Extracting the $s_\tau=0$ term, we generate this expression as 
\begin{equation}
    H_{i\leftarrow \tau } (z) = (1-\phi)^2 + \sum_{s_\tau=1}^\infty z^{s_\tau} \pi_{i\leftarrow\tau}(s_\tau).\label{eq:Hitau}
\end{equation}
Summing over $s_\tau$, expressions \ref{eq:subtj} and \ref{eq:subtk} follow similar manipulations to Eq \ref{eq:zlogic}; whilst Eq \ref{eq:double} becomes
\begin{align}
\sum_{s_\tau=1}^\infty z^{s_\tau}\pi_{i\leftarrow \tau}\left(s_\tau\ \big|\ 
    \begin{tikzpicture}[baseline=-0.25ex,scale=0.5]
    \node[MyVertex] (middle) at ( 0,   -0.2) {};
    \node[MyVertex] (left)   at (-0.5, 0.65) {};
    \node[MyVertex] (right)  at ( 0.5, 0.65) {};
    \draw[MyFullEdge,black]
          (middle) -- (left);
    \draw[MyFullEdge,black]
          (middle) -- (right);
    \draw[MyDashedEdge,black]
          (left) -- (right);
\end{tikzpicture},    \begin{tikzpicture}[baseline=-0.25ex,scale=0.5]
    \node[MyVertex] (middle) at ( 0,   -0.2) {};
    \node[MyVertex] (left)   at (-0.5, 0.65) {};
    \node[MyVertex] (right)  at ( 0.5, 0.65) {};
    \draw[MyFullEdge,black]
          (middle) -- (left);
    \draw[MyDashedEdge,black]
          (middle) -- (right);
    \draw[MyFullEdge,black]
          (left) -- (right);
\end{tikzpicture},
\begin{tikzpicture}[baseline=-0.25ex,scale=0.5]
    \node[MyVertex] (middle) at ( 0,   -0.2) {};
    \node[MyVertex] (left)   at (-0.5, 0.65) {};
    \node[MyVertex] (right)  at ( 0.5, 0.65) {};
    \draw[MyDashedEdge,black]
          (middle) -- (left);
    \draw[MyFullEdge,black]
          (middle) -- (right);
    \draw[MyFullEdge,black]
          (left) -- (right);
\end{tikzpicture},
\begin{tikzpicture}[baseline=-0.25ex,scale=0.5]
    \node[MyVertex] (middle) at ( 0,   -0.2) {};
    \node[MyVertex] (left)   at (-0.5, 0.65) {};
    \node[MyVertex] (right)  at ( 0.5, 0.65) {};
    \draw[MyFullEdge,black]
          (middle) -- (left);
    \draw[MyFullEdge,black]
          (middle) -- (right);
    \draw[MyFullEdge,black]
          (left) -- (right);
\end{tikzpicture}
\right) =&\ \sum_{s_\tau=1}^\infty z^{s_\tau} \sum_{\{s_\ell:\ell\in \partial_\nu( j ),\ \nu \in \bm \nu_{ j }\backslash \tau\}} \sum_{\{s_r:r\in \partial_\varphi( k ),\ \varphi \in \bm \varphi_{ k }\backslash \tau\}}\nonumber\\
&\ \times\left[\prod_{\nu\in \bm \nu_{ j }\backslash \tau} \prod_{\ell\in\partial_\nu( j )}\pi_{ j \leftarrow \ell}(s_\ell)\prod_{\varphi\in \bm \varphi_{ k }\backslash \tau} \prod_{r\in\partial_\varphi( k )}\pi_{ k \leftarrow r}(s_r)\right]\nonumber\\
&\ \times \delta\left(s_\tau-2,\sum_{\nu\in \bm \nu_{ j }\backslash \tau} \sum_{\ell\in\partial_\nu( j )}s_\ell+
\sum_{\varphi\in \bm \varphi_{ k }\backslash \tau} \sum_{r\in\partial_\varphi( k )}s_r
\right).
\end{align}
The Kronecker delta evaluates as follows
\begin{align}
    \sum^\infty_{s_\tau=1}z^{s_\tau}\stackrel{\delta}{=}&\ z^2\sum^\infty_{s=1}z^{\sum\limits_{\nu\in \bm \nu_{ j }\backslash \tau} \sum\limits_{\ell\in\partial_\nu( j )}s_\ell}z^{
\sum\limits_{\varphi\in \bm \varphi_{ k }\backslash \tau} \sum\limits_{r\in\partial_\varphi( k )}s_r},\\
=&\  z^2\prod_{\nu\in \bm \nu_{ j }\backslash \tau}\prod_{\ell\in\partial_\nu( j )}
\sum^\infty_{s_\ell=0}
\prod_{\varphi\in \bm \varphi_{ k }\backslash \tau}\prod_{r\in\partial_\varphi( k )}\sum^\infty_{s_r=0}z^{s_\ell+s_r}.
\label{eq:zlogic2}
\end{align}
We find
\begin{align}
\sum_{s_\tau=1}^\infty z^{s_\tau}\pi_{i\leftarrow \tau}\left(s_\tau\ \big|\ 
    \begin{tikzpicture}[baseline=-0.25ex,scale=0.5]
    \node[MyVertex] (middle) at ( 0,   -0.2) {};
    \node[MyVertex] (left)   at (-0.5, 0.65) {};
    \node[MyVertex] (right)  at ( 0.5, 0.65) {};
    \draw[MyFullEdge,black]
          (middle) -- (left);
    \draw[MyFullEdge,black]
          (middle) -- (right);
    \draw[MyDashedEdge,black]
          (left) -- (right);
\end{tikzpicture},    \begin{tikzpicture}[baseline=-0.25ex,scale=0.5]
    \node[MyVertex] (middle) at ( 0,   -0.2) {};
    \node[MyVertex] (left)   at (-0.5, 0.65) {};
    \node[MyVertex] (right)  at ( 0.5, 0.65) {};
    \draw[MyFullEdge,black]
          (middle) -- (left);
    \draw[MyDashedEdge,black]
          (middle) -- (right);
    \draw[MyFullEdge,black]
          (left) -- (right);
\end{tikzpicture},
\begin{tikzpicture}[baseline=-0.25ex,scale=0.5]
    \node[MyVertex] (middle) at ( 0,   -0.2) {};
    \node[MyVertex] (left)   at (-0.5, 0.65) {};
    \node[MyVertex] (right)  at ( 0.5, 0.65) {};
    \draw[MyDashedEdge,black]
          (middle) -- (left);
    \draw[MyFullEdge,black]
          (middle) -- (right);
    \draw[MyFullEdge,black]
          (left) -- (right);
\end{tikzpicture},
\begin{tikzpicture}[baseline=-0.25ex,scale=0.5]
    \node[MyVertex] (middle) at ( 0,   -0.2) {};
    \node[MyVertex] (left)   at (-0.5, 0.65) {};
    \node[MyVertex] (right)  at ( 0.5, 0.65) {};
    \draw[MyFullEdge,black]
          (middle) -- (left);
    \draw[MyFullEdge,black]
          (middle) -- (right);
    \draw[MyFullEdge,black]
          (left) -- (right);
\end{tikzpicture}
\right) =&\ 
    \bigg(z\prod_{\nu\in \bm \nu_{ j }\backslash \tau}\prod_{\ell\in\partial_\nu( j )}\sum^\infty_{s_\ell=0}\pi_{ j \leftarrow \ell}(s_\ell)z^{s_\ell}\bigg)\bigg(z
\prod_{\varphi\in \bm \varphi_{ k }\backslash \tau}\prod_{r\in\partial_\varphi( k )}\sum^\infty_{s_r=0}\pi_{ k \leftarrow r}(s_r)z^{s_r}\bigg),\\
=&\  \bigg(z\prod_{\nu\in \bm \nu_{ j }\backslash \tau}H_{ j \leftarrow \nu}(z)\bigg)\bigg(z
\prod_{\varphi\in \bm \varphi_{ k }\backslash \tau}H_{ k \leftarrow \varphi}(z)\bigg).
\end{align}
Inserting this result into Eq \ref{eq:Hitau} together with the other expressions in Eqs \ref{eq:subtj} and \ref{eq:subtk} we have
\begin{align}
    H_{i\leftarrow \tau}(z) =&\  (1-\phi)^2 + \phi(1-\phi)^2\left(z\prod_{\nu\in \bm \nu_{ j }\backslash \tau}H_{ j \leftarrow \nu}(z) + z\prod_{\varphi\in \bm \varphi_{ k }\backslash \tau}H_{ k \leftarrow \varphi}(z)\right)\nonumber\\
    &\ +\left(3\phi^2(1-\phi)+\phi^3 \right)
    \bigg(z\prod_{\nu\in \bm \nu_{ j }\backslash \tau}H_{ j \leftarrow \nu}(z)\bigg)\bigg(z
\prod_{\varphi\in \bm \varphi_{ k }\backslash \tau}H_{ k \leftarrow \varphi}(z)\bigg).\label{eq:triangle_Final}
\end{align}
The result is a polynomial in powers of generating functions $H_{\ell\leftarrow \sigma}(z)$ where $\ell$ is a vertex in motif $\tau$ and $\sigma$ is a motif that $\ell$ belongs to, other than motif $\tau$.

\subsection{Calculating $H_{i\leftarrow \tau}$ for chordless cycles of size $n\geq 3$}
\label{sec:cycles_formula}
Perhaps the simplest family of subgraphs to consider is the set of chordless cycles of increasing vertex count. In this section we will derive a closed form expression for $H_{i\leftarrow \tau}$ where $\tau$ is a cycle of length $n\geq 3$. As with the triangle in section \ref{subsec:3-clique} we choose a vertex in the cycle $i$ and calculate the probability that $s_\tau$ other vertices can be reached due to membership in motif $\tau$. We achieve this by writing the conditional probabilities that the cycle has a given bond configuration; since, each bond configuration occurs with a different probability. Let all edges of the cycle be occupied and so vertex $i$ belongs to the same component as the other $n-1$ vertices. Without loss of generality, let us also label the vertices from $1$ to $n$ and set $i=1$. In this case we can write 
\begin{align}
    \pi_{i\leftarrow \tau}(s_\tau\mid \mathcal C_n)=\sum_{\{s_\ell:\ell\in \partial_\nu( j ),\ \nu \in \bm \nu_{ j }\backslash \tau,\  j \in \tau\backslash i\}} \left[ \prod_{ j \in \tau\backslash i}\prod_{\nu\in \bm \nu_{ j }\backslash \tau}\prod_{\ell\in\partial _\nu( j )} \pi_{ j \leftarrow \ell}(s_\ell) \right]\delta\left(s_\tau-n+1,  \sum_{ j \in \tau\backslash i}\sum_{\nu\in \bm \nu_{ j }\backslash \tau}\sum_{\ell\in\partial _\nu( j )}s_\ell\right)\label{eq:cycletogether}.
    \end{align}
The set ${\{\ell\in \partial_\nu( j ),\ \nu \in \bm \nu_{ j }\backslash \tau,\  j \in \tau\backslash i\}} $ accounts (from left to right) for all edges $\ell$ in motif $\nu$ that vertex $ j $ belongs to, for all motifs $\nu$ that vertex $ j $ belongs to, apart from $\tau$, for all vertices $ j $ in motif $\tau$ apart from $i$. The Kronecker delta filters those terms that do not sum to $s_\tau-n+1$, accounting for the vertices that belong to the $n$-cycle other than $i$. If one of the edges around the cycle was set to the unoccupied state a path between the vertices through the cycle would still be present. Subsequent removal of an edge from the cycle has the potential to isolate a vertex and so, all of the states with a single edge removed have been exhausted. We can write the coefficient of this term as 
\begin{equation}
    P(\mathcal C_n)=\phi^{n} + n\phi^{n-1}(1-\phi).
\end{equation}
We now examine the configuration $\mathcal C_{n-1}$ where a single neighbour vertex $ k $ within the cycle belongs to a different component to vertex $i$. For this to occur, both of the edges that connect this vertex to the cycle must be unoccupied with probability $(1-\phi)^2$ and so the probability of this edge configuration is 
\begin{equation}
    P(\mathcal C_{n-1}) = \phi^{n-1-1}(1-\phi)^2.
\end{equation}
To exclude $ k $ from the set of vertices in motif $\tau$ that contribute to the summation in Eq \ref{eq:cycletogether}, we simply modify the set notation to $\mathcal S( k )={\{\ell\in \partial_\nu( j ),\ \nu \in \bm \nu_{ j }\backslash \tau, \ j \in \tau\backslash\{ i, k \}\}} $. However, given that the identity of $ k $ can be any neighbour in $\tau$ apart from $i$, we must account for all possible identities by summing $ k $ from $2$ to $n$ to obtain
\begin{align}
    \pi_{i\leftarrow \tau}(s_\tau\mid \mathcal C_{n-1})=\sum_{ k }\sum_{\mathcal S( k )}\left[ \prod_{ j \in \tau\backslash \{i, k \}}\prod_{\nu\in \bm \nu_{ j }\backslash \tau}\prod_{\ell\in\partial _\nu( j )} \pi_{ j \leftarrow \ell}(s_\ell) \right]\delta\left(s_\tau-n+2,  \sum_{ j \in \tau\backslash \{i, k \}}\sum_{\nu\in \bm \nu_{ j }\backslash \tau}\sum_{\ell\in\partial _\nu( j )}s_\ell\right)\label{eq:cycletogether2}.
\end{align}
No edges can be removed from this motif without further isolation of a vertex, indicating that this edge configuration has been fully calculated. Considering the next term $\mathcal C_{n-2}$, where the removal of another edge isolates a second vertex. Vertex $i$ now belongs to an induced subgraph of $n-2$ vertices in $\tau$. We cannot set an edge of our choice to be unoccupied to arrive at this state, however. It happens that the removed edge \textit{must} be one of the edges that a neighbour of $ k $ has within the component of vertex $i$. If we had chosen a different edge, we would not account for all combinations for connected components of this length as we would have isolated vertices prematurely from the chain of removed vertices. We can write 
\begin{align}
    \pi_{i\leftarrow \tau}(s_\tau\mid \mathcal C_{n-2})=\sum_{ k =2}^{{n-1}}\sum_{\mathcal S( k ,\tau_{k+1})}\left[ \prod_{ j \in \tau\backslash \{i, k ,\tau_{k+1}\}}\prod_{\nu\in \bm \nu_{ j }\backslash \tau}\prod_{\ell\in\partial _\nu( j )} \pi_{ j \leftarrow \ell}(s_\ell) \right]\delta\left(s_\tau-n+3,  \sum_{ j \in \tau\backslash \{i, k ,\tau_{k+1}\}}\sum_{\nu\in \bm \nu_{ j }\backslash \tau}\sum_{\ell\in\partial _\nu( j )}s_\ell\right)\label{eq:cycletogether3},
\end{align}
where $\mathcal S( k ,\tau_{k+1}) = {\{\ell\in \partial_\nu( j ),\ \nu \in \bm \nu_{ j }\backslash \tau,  \ j \in \tau\backslash\{ i, k ,\tau_{k+1}\}\}}$. To picture this, we are essentially moving a pair of connected vertices around the motif as a sliding window. The probability of this edge configuration is 
\begin{equation}
    P(\mathcal C_{n-2}) = \phi^{n-1-2}(1-\phi)^2.
\end{equation}
The logic behind this expression is that the chain of removed vertices must fail to connect to the same component that vertex $i$ belongs to, which occurs with probability $(1-\phi)^2$. Otherwise, all edges between the vertices that are connected to $i$ are occupied with probability $\phi$, of which, there are $n-1-2$. 

Generalising this process, the conditional probability that a chordless cycle with $n$ vertices, of which $r$ belong to a different component to vertex $i=1$, leads to $s_\tau$ reachable vertices is 
\begin{align}
     \pi_{i\leftarrow \tau}(s_\tau\mid \mathcal C_{n-r})P(\mathcal C_{n-r}) =&\  \phi^{n-1-r}(1-\phi)^2\sum_{ k =2}^{n-r+1}\sum_{\mathcal S([ k ,\tau_{k+r-1}])}\left[ \prod_{ j \in \tau\backslash \{i,[ k ,\tau_{k+r-1}]\}}\prod_{\nu\in \bm \nu_{ j }\backslash \tau}\prod_{\ell\in\partial _\nu( j )} \pi_{ j \leftarrow \ell}(s_\ell) \right]\nonumber\\
     &\ \times\delta\left(s_\tau-(n-1-r),  \sum_{ j \in \tau\backslash \{i,[ k ,\tau_{k+r-1}]\}}\sum_{\nu\in \bm \nu_{ j }\backslash \tau}\sum_{\ell\in\partial _\nu( j )}s_\ell\right)\label{eq:cycletogether4},
\end{align}
where there are $n-1-r$ vertices in $\tau$ that are connected to $i$ and with the notational understanding that all vertices in the range $ k $ to $\tau_{k+r-1}$ (inclusive) are excluded from the set in addition to vertex $i$ such that  
\begin{equation}
     \mathcal S([ k ,\tau_{k+r-1}]) = {\{\ell\in \partial_\nu( j ),\ \nu \in \bm \nu_{ j }\backslash \tau, \ j \in \tau\backslash\{ i, k ,\tau_{k+1},\dots,\tau_{k+r-2},\tau_{k+r-1}\}\}},
\end{equation}
and similarly for the summation in the Kronecker delta 
\begin{equation}
     j \in \tau\backslash \{i,[ k ,\tau_{k+r-1}]\} =  { j \in \tau\backslash\{ i, k ,\tau_{k+1},\dots,\tau_{k+r-2},\tau_{k+r-1}\}}.
\end{equation}
We can then sum Eq \ref{eq:cycletogether4} from $r=1$ to $n-1$ to account for chains of isolated vertices for all lengths. Together with the expression for the fully connected cycle we finalise the expression 
\begin{equation}
    \pi_{i\leftarrow \tau}(s_\tau) = \pi_{i\leftarrow \tau}(s_\tau\mid \mathcal C_n)P(\mathcal C_n) + \sum_{r=1}^{n-1} \pi_{i\leftarrow \tau}(s_\tau\mid \mathcal C_{n-r})P(\mathcal C_{n-r}) .
\end{equation}
This expression is then generated as 
\begin{align}
    H_{i\leftarrow \tau}(z) =&\ \left(\phi^{n} + n\phi^{n-1}(1-\phi)\right)\prod_{ j \in \tau\backslash i}\left(z\prod_{\nu\in \bm \nu_{ j }\backslash\tau} H_{ j \leftarrow \nu}(z)\right)\nonumber\\
    &\ +\sum_{r=1}^{n-1} \phi^{n-1-r}(1-\phi)^2 \sum_{ k =2}^{n-r+1}\prod_{ j \in\tau\backslash\{i,[ k ,\tau_{k+r-1}]\}}\left(z\prod_{\nu\in \bm\nu_{ j }\backslash \tau} H_{ j \leftarrow \nu}(z)\right).\label{eq:main}
\end{align}

The first term in Eq \ref{eq:main} accounts for the edge configuration when all vertices in the $n$-cycle are connected.
The second term in this expression accounts for the various edge configurations that the cycle may exhibit when a chain of $r$ vertices are removed from the cycle due to bond percolation. From right to left the indices of the second term account for: for all motifs $\nu$ (apart from $\tau$) that vertex $ j $ belongs to, for all $ j $ in the chain of vertices that are connected to $i$ (which we enumerate by removing vertices $\{i,[ k ,\tau_{k+r-1}]\}$ from the set of vertices in $\tau$), for all possible starting points $ k $ of the $r$-chain, for all possible lengths $r$. This expression is the main result of this section; as a consistency check, we find Eq \ref{eq:triangle_Final} for triangles $n=3$ can be recovered.

\section{Calculating $H_{i\leftarrow \tau}$ for cliques}
\label{sec:cliques_formula}

We will now derive an exact expression for belief propagation on clique motifs, extending the ensemble expressions derived by Mann \textit{et al} \cite{PhysRevE.104.024304}. The generating function for the probability that a focal vertex $i$, that is member of a clique $\tau$, can reach $s_\tau$ vertices through its membership in $\tau$ is $H_{i\leftarrow \tau}(z)$. Following bond percolation, each edge in $\tau$ can be occupied with probability $\phi$ or unoccupied with probability $1-\phi$. Depending on the bond configuration of the clique, vertex $i$ might not be connected to all neighbours via occupied edges. To calculate $H_{i\leftarrow \tau}(z)$ we must average over each possible bond configuration that the focal vertex might observe and the various states of connectivity among the neighbours and $i$. The expression takes the form of a polynomial whose terms are the powerset, denoted by $2^{\{ j \in \tau\backslash i\}}$, of the set of vertices in the clique apart from $i$, including the empty set $\{\varnothing\}$. For instance, consider the form of the expression when $\tau$ is a $4$-clique, with vertices labelled 0 to 3 and with vertex $i$ arbitrarily being labelled as 0, we have
\begin{equation}
    H_{0\leftarrow \tau}(z) = \mathcal P\left(\varnothing\right)+ \mathcal P\left(1\right)
    + \mathcal P\left(2\right)+ \mathcal P\left(3\right)+ \mathcal P\left(1,2\right)+ \mathcal P\left(1,3\right)+ \mathcal P\left(2,3\right) + \mathcal P\left(1,2,3\right).\label{eq:polynomial}
\end{equation}
Each term $\mathcal P( j ,\dots,  k )$ generates the probability that membership in $\tau$ leads to $s_\tau$ reachable vertices, given $i$ is connected to vertices $ j ,\dots,  k $ (a path of occupied edges exits within the motif). Each term can be decomposed into the probability that the neighbours collectively lead to $s_\tau$ vertices (including themselves), multiplied by the probability that the bond percolation process yielded that edge configuration $P(j,\dots,k)$
\begin{equation}
    \mathcal P( j ,\dots, k ) = \sum_{s_\tau=1}^\infty z^{s_\tau}\pi_{i\leftarrow \tau} (s_\tau\mid  j ,\dots, k ) P( j ,\dots, k ).
\end{equation}
It happens, that because of the symmetry of a clique, $P( j ,\dots, k )$ only depends on the number of vertices $| j ,\dots, k |$, not their identity \footnote{This is certainly not true for motifs whose vertices do not belong to the same site (or more correctly - orbit). In that case, the bond occupancy probability depends on the identity of the vertices in the connected component (see \cite{PhysRevE.104.024304}).} Given this logic, we group terms that have equal numbers of connected vertices in $\tau$ and the polynomial in Eq \ref{eq:polynomial} becomes a summation over the subsets $a_\kappa\in A_\kappa$ of the set of neighbours $\{ j \in \tau\backslash i\}$ of a given length $\kappa$
\begin{equation}
    H_{i\leftarrow \tau}(z) = \sum_{\kappa=0}^{|\tau|-1}P(\kappa)\sum_{a_{\kappa}\in A_{\kappa}}\sum_{s_\tau=0}^{\infty}z^{s_\tau}\pi_{i\leftarrow \tau} (s_\tau\mid a_\kappa),\label{eq:main_clique_undone}
\end{equation}
Concentrating on the probability $\pi_{i\leftarrow \tau}(s_\tau\mid a_\kappa)$ first, we can expand this as a sum over all sizes $s_\ell$
\begin{equation}
    \pi_{i\leftarrow \tau} (s_\tau\mid a_\kappa) =\sum_{\{s_\ell:\ell\in\partial_\nu( j ),\ \nu\in \bm \nu_{ j }\backslash \tau,\ j \in a_\kappa\}} \left[\prod_{ j \in a_\kappa}\prod_{\nu\in\bm\nu_{ j }\backslash\tau}\prod_{\ell\in\partial_\nu( j )}\pi_{ j \leftarrow\ell}(s_{\ell})\right]\delta\left(s_\tau - \kappa -1,
    \sum_{ j \in a_\kappa}\sum_{\nu\in\bm\nu_{ j }\backslash\tau}\sum_{\ell\in\partial_\nu( j )}s_\ell\right).
\end{equation}
where $\ell$ is an index over the edges of the neighbours that $i$ is connected to in $\tau$, excluding those that point back to $\tau$. Generating this expression we obtain 
\begin{equation}
    \sum_{s_\tau=0}^{\infty}z^{s_\tau}\pi_{i\leftarrow \tau}(s_\tau\mid a_\kappa) = \prod_{ j \in a_\kappa}\left(z\prod_{\nu\in\bm \nu_{ j }\backslash\tau} H_{ j \leftarrow\nu}(z)\right).\label{eq:cliquepi}
\end{equation}
The next step is to find the coefficient $P(\kappa)$ for each size $\kappa$ component of neighbours within $\tau$ that can attach to $i$, see \cite{PhysRevE.104.024304}. Within this calculation, we must account for all possible connected graphs that can occur among $\kappa+1$ vertices in $\tau$. For a given  $\kappa$ there are $\frac 12(\kappa+1)\kappa$ total edges among vertices in the connected component. Letting $r=|\tau|-\kappa-1$, there are $\frac 12 r(r-1)$ many edges between the removed vertices (ones that are not in the component with $i$). Finally, there are 
\begin{equation}
    \omega(r) = \sum^r_{v=1}(|\tau|-v) -\frac 12 r(r-1),
\end{equation}
edges that connect vertices that belong to the same component as $i$ to those that don't (they interface the two components).

\begin{figure}[ht!]
    \centering
    \includegraphics[width=0.25\textwidth]{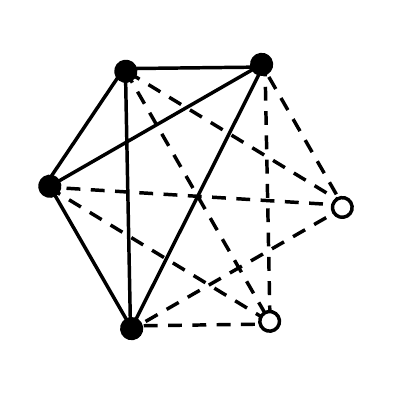}
    \caption{A bond occupancy configuration of a 6-clique in which two vertices do not belong to the giant component (unfilled) whilst the remaining 4 vertices do. Solid edges are occupied whilst dashed edges are unoccupied. The occupation state of the edge linking the two unfilled vertices is inconsequential to the percolation properties of the 4 vertices in the giant component. There are $0.5(3+1)3=6$ occupied edges among the filled vertices, of which 3 can be set to unoccupied and connectivity retained. There are $\omega(2)=8$ edges that must be unoccupied if the two unfilled vertices are to remain outside of the giant component.
    }
    \label{fig:6clique}
\end{figure}

For instance, consider a $|\tau|=6$ clique that has $\kappa=3$ neighbours that are connected to focal vertex $i$ and therefore $r=2$ vertices that do not have a path of occupied edges to $i$ as depicted in Fig \ref{fig:6clique}. There are $\frac 12(3+1)3=6$ edges among the vertices in the same component as $i$ and $\omega(2)=8$ interface edges. The bond occupancy probability of this configuration is given by $\phi^{\frac 12(\kappa+1)\kappa}(1-\phi)^{\omega(r)}$. However, we can remove up to $\frac 12\kappa(\kappa-1)$ of the $\frac 12(\kappa+1)\kappa$ occupied edges from the component that $i$ belongs to and still retain connectivity among the $\kappa+1$ vertices. Therefore, indexing the number of removed edges by $m$ we have 
\begin{equation}
    P(\kappa) = \sum_{m=0}^{\frac 12 \kappa(\kappa-1)}\mathcal Q_{\kappa+1,\frac 12(\kappa+1)\kappa-m}\phi^{\frac 12(\kappa+1)\kappa-m}(1-\phi)^{\omega(r)+m},\label{eq:pkclique}
\end{equation}
where $\mathcal Q_{n,k}$ is the number of connected graphs that can be made that have $n$ vertices and $k$ edges, see Appendix \ref{sec:appendix:connected_graphs}. Inserting Eqs \ref{eq:cliquepi} and \ref{eq:pkclique} into Eq \ref{eq:main_clique_undone} we arrive at the main result of this section
\begin{equation}
    H_{i\leftarrow \tau}(z) = \sum_{\kappa=0}^{|\tau|-1}\sum_{m=0}^{\frac 12 \kappa(\kappa-1)}\mathcal Q_{\kappa+1,\frac 12(\kappa+1)\kappa-m}\phi^{\frac 12(\kappa+1)\kappa-m}(1-\phi)^{\omega(r)+m}\sum_{a_{\kappa}\in A_{\kappa}}\prod_{ j \in a_\kappa}\left(z\prod_{\nu\in\bm \nu_{ j }\backslash\tau} H_{ j \leftarrow\nu}(z)\right),\label{eq:main_clique}
\end{equation}
where  $r=|\tau|-\kappa-1$. Unpacking this expression when $\tau$ is a 3-clique reproduces Eq \ref{eq:triangle_Final}. When $\tau$ is a 4-clique we obtain
\begin{align}
    H_{i\leftarrow \tau} (z)=&\ (1-\phi)^3 + z\phi(1-\phi)^4\left[\prod_{\nu\in\bm \nu_{\tau_1}\backslash\tau} H_{\tau_1\leftarrow\nu} (z)+\prod_{\nu\in\bm \nu_{\tau_2}\backslash\tau} H_{\tau_2\leftarrow\nu} (z)+\prod_{\nu\in\bm \nu_{\tau_3}\backslash\tau} H_{\tau_3\leftarrow\nu} (z)\right]\nonumber\\
    & +z^2\left[\phi^3(1-\phi)^3 + 3\phi^2(1-\phi)^4\right]\left[\prod_{\nu\in\bm \nu_{\tau_1}\backslash\tau} H_{\tau_1\leftarrow\nu} (z)\prod_{\nu\in\bm \nu_{\tau_2}\backslash\tau} H_{\tau_2\leftarrow\nu}  (z)+ \prod_{\nu\in\bm \nu_{\tau_1}\backslash\tau} H_{\tau_1\leftarrow\nu} (z)\prod_{\nu\in\bm \nu_{\tau_3}\backslash\tau} H_{\tau_3\leftarrow\nu} (z)\right.\nonumber\\
    &+\left. \prod_{\nu\in\bm \nu_{\tau_2}\backslash\tau} H_{\tau_2\leftarrow\nu}(z)\prod_{\nu\in\bm \nu_{\tau_3}\backslash\tau} H_{\tau_3\leftarrow\nu}(z)\right]\nonumber\\
    &+z^3 [\phi^6+6\phi^5(1-\phi) + 15\phi^4(1-\phi)^2+16\phi^3(1-\phi)^3]\prod_{\nu\in\bm \nu_{\tau_1}\backslash\tau} H_{\tau_1\leftarrow\nu}(z)\prod_{\nu\in\bm \nu_{\tau_2}\backslash\tau} H_{\tau_2\leftarrow\nu}(z)\prod_{\nu\in\bm \nu_{\tau_3}\backslash\tau} H_{\tau_3\leftarrow\nu}(z).
\end{align}
The coefficients of larger cliques are in agreement with the exact expressions previously found by \cite{PhysRevE.68.026121} and \cite{PhysRevE.104.024304}. The derivative of Eq \ref{eq:main_clique} that is required for the calculation of the finite components is given by 
\begin{equation}
    H'_{i\leftarrow \tau}(z) = \sum_{\kappa=0}^{|\tau|-1}\sum_{m=0}^{\frac 12 \kappa(\kappa-1)}\mathcal Q_{\kappa+1,\frac 12(\kappa+1)\kappa-m}\phi^{\frac 12(\kappa+1)\kappa-m}(1-\phi)^{\omega(r)+m}\sum_{a_{\kappa}\in A_{\kappa}}\frac{d}{dz}\left[\prod_{ j \in a_\kappa}\left(z\prod_{\nu\in\bm \nu_{ j }\backslash\tau} H_{ j \leftarrow\nu}(z)\right)\right],\label{eq:main_clique_derivative}
\end{equation}
where we can apply Eq \ref{eq:derivative1} to find
\begin{align}
    \frac{d}{dz}\left[\prod_{ j }\left(z\prod_\nu H_{{ j }\leftarrow \nu}(z)\right)\right] =&\  \prod_{ j }\left(z\prod_\nu H_{{ j }\leftarrow \nu}(z)\right)\left[\sum_{ j }\left(z\prod_\nu H_{{ j }\leftarrow \nu}(z)\right)^{-1}\right.\nonumber\\
    &\left.\ \times\left[\prod_\nu H_{{ j }\leftarrow \nu}(z) + z\left(\prod_\nu 
H_{{ j }\leftarrow \nu }(z)\right)\left(\sum_\nu \frac{H_{{ j }\leftarrow \nu }'(z)}{H_{{ j }\leftarrow \nu }(z)}\right)\right]\right].
\end{align}
The fixed point of these expressions can be found to yield the derivatives and Eq \ref{eq:small_component} can be solved.

\end{widetext}

\section{Relation to the generalised configuration model}\label{sec:configuration_model}

The message passing formulation calculates the properties of a given graph realisation $G$. Often, the properties of an \textit{ensemble} of networks $G\in\mathcal G$ with equivalent statistics are required, rather than a single instance. An ensemble of random networks can be created according to the generalised configuration model algorithm \cite{karrer_newman_2010, PhysRevLett.103.058701,PhysRevE.80.020901,PhysRevE.103.012313,PhysRevE.103.012309}. In this model, a set of motifs are defined and each vertex is assigned a tuple of integers, called its \textit{joint degree} that represents the number of edge-disjoint motifs of a given topology that it belongs to. For instance, if a vertex belongs to three ordinary edges, one triangle, one 4-cycle and two 5-cycles the joint degree is $(k_2,k_3,k_4,k_5)=(3,1,1,2)$. The distribution of joint degrees are fixed by the joint degree distribution $p(k_2,k_3,k_4,k_5)$ which is the fraction of vertices with a given joint degree. 

During the construction process, the configuration model randomises the identity of the vertices that belong to a given motif. Therefore, for a given vertex $i$, the joint degree of its neighbours can vary drastically and so the neighbourhood beyond the nearest neighbours becomes a mean field quantity. To see this, consider an edge-disjoint triangle cover of a network; ordinary edges being present also. All of the edges in the graph have been assigned to a \textit{motif topology} by the cover that relates to the type of motif to which they belong. Let us now break each edge into two parts whilst retaining the topology labels; isolating each vertex. To create the required number of ordinary edges and triangles, the configuration model selects vertices at random and connects their stubs together, matching the topologies. For instance, to create a triangle, three vertices are selected at random that have free unmatched triangle stubs and are then connected together appropriately. In this way, each random graph that is constructed by the configuration model belongs to an equivalent set of motif topologies, but the identity of the vertices that comprise those motifs will be different. Representing this stochasticity by an average over all graphs in the ensemble, the product of the average message that vertex $ j \in \tau$ receives from each of its other motifs $\nu\in\bm \nu_{ j }\backslash\tau$ can be written as an average over the product of messages 
\begin{equation}
      \prod_{\nu\in\bm \nu_{ j }\backslash \tau}\langle H_{\nu}(z)\rangle\leq\left\langle z\prod_{\nu\in\bm\nu_{ j }\backslash \tau}H_{ j \leftarrow \nu}(z)\right\rangle,\label{eq:ave_prod}
\end{equation}
where we have used a version of the Chebyshev integral inequality \cite{PhysRevE.82.016101} for $k$ monotonic functions of the same monotonicity
\begin{equation}
   \prod^n_{i=1}\left\langle f_i(x_1,\dots,x_k)\right\rangle\leq  \left\langle \prod_{i=1}^nf_i(x_1,\dots,d_k)\right\rangle.
\end{equation}
In general, the average of a product is not equal to the product of the average. Only when the covariance between the messages is zero is this true. For any two motifs to be correlated in the cavity graph, there must be loops in the factor graph. However, in the limit that the number of vertices $N$ becomes infinite, the shortest short range cycles are expected to be at least length $\mathcal O(\log N)$. Assuming that the factor graph is locally treelike, then the messages that arrive at a cavity from independent motifs can be treated as though they are uncorrelated with one another; in this case Eq \ref{eq:ave_prod} becomes an equality. 

As an example, let us partition the product over the motifs of the neighbour into a product over its motif topologies. For instance, for ordinary edges and triangles we have
\begin{equation}
  \prod_{s\in\bm s_{ j }\backslash \tau}\langle H_{\bot}\rangle\prod_{t\in\bm t_{ j }}\langle H_{\Delta}\rangle =\left\langle z\prod_{s\in\bm s_{ j }\backslash \tau}H_{ j \leftarrow s}\prod_{t\in\bm t_{ j }}H_{ j \leftarrow t}\right\rangle,
\end{equation}
and
\begin{equation}
     \prod_{s\in\bm s_{ j }}\langle H_{\bot}\rangle\prod_{t\in\bm t_{ j }\backslash \tau}\langle H_{\Delta}\rangle=\prod_{ j \in t\backslash i}\left\langle z\prod_{s\in\bm s_{ j }}H_{ j \leftarrow s}\prod_{t\in\bm t_{ j }\backslash t}H_{ j \leftarrow t}\right\rangle,
\end{equation}
where $\bm s_{ j }$ $(\bm t_{ j })$ is the set of ordinary edges (triangles) to which vertex $ j $ belongs. There are two expressions since motif $\tau$ could have been an ordinary edge or a triangle and the probabilities associated with each one are not equivalent in general. However, since the average message is the same for each motif of a given topology, we can simplify this as a power
\begin{equation}
     \prod_{s\in\bm s_{ j }\backslash \tau}\langle H_{\bot}\rangle\prod_{t\in\bm t_{ j }}\langle H_{\Delta}\rangle = \langle H_{\bot}\rangle^{k_2}\langle H_{\Delta}\rangle^{k_3},
\end{equation}
where $k_2$ is the excess number of ordinary edges that $ j $ has (given that $\tau$ is an ordinary edge) and $k_3$ is the number of triangles $ j $ belongs to. In other words, $k_2=\text{card} (\bm s_{ j }\backslash \tau)$ and $k_3=\text{card}(\bm t_{ j })$. Similarly, if $\tau$ had instead been a triangle we can write
\begin{equation}
     \prod_{s\in\bm s_{ j }}\langle H_{\bot,s}\rangle\prod_{t\in\bm t_{ j }\backslash \tau}\langle H_{\bot,t}\rangle  = \langle H_{\bot}\rangle^{k_2}\langle H_{\Delta}\rangle^{k_3},
\end{equation}
where $k_2=\text{card}(\bm s_{ j })$ is the number of ordinary edges and $k_3=\text{card}(\bm t_{ j }\backslash \tau)$ is the excess triangle degree of vertex $ j $, respectively. This expression is the product of the average messages along ordinary edges and triangles. The excess degree distributions are distributed as $q_\tau(k_2,k_3)$ and so averaging over the distribution we can write a generating function for the message that a neighbour receives in the random graph ensemble as
\begin{align}
    G_{1,\tau} (z)= \sum_{k_2}\sum_{k_3}q_\tau \langle H_{\bot}(z)\rangle^{k_2}\langle H_{\Delta}(z)\rangle^{k_3}.
\end{align}
An equivalent argument can be followed to define another fundamental generating function $G_0(x,y)=\sum_{k_2}\sum_{k_3}p(k_2,k_3)x^{k_2}y^{k_3}$ from Eq \ref{eq:G_i}. This logic can be extended to all motifs that are included in the model. With these two generating functions defined, the mapping between the message passing formulation and the generalised configuration model is complete. 

\section{Discussion}
\label{sec:discussion}

\begin{figure}[ht!]
    \centering
    \includegraphics[width=0.5\textwidth]{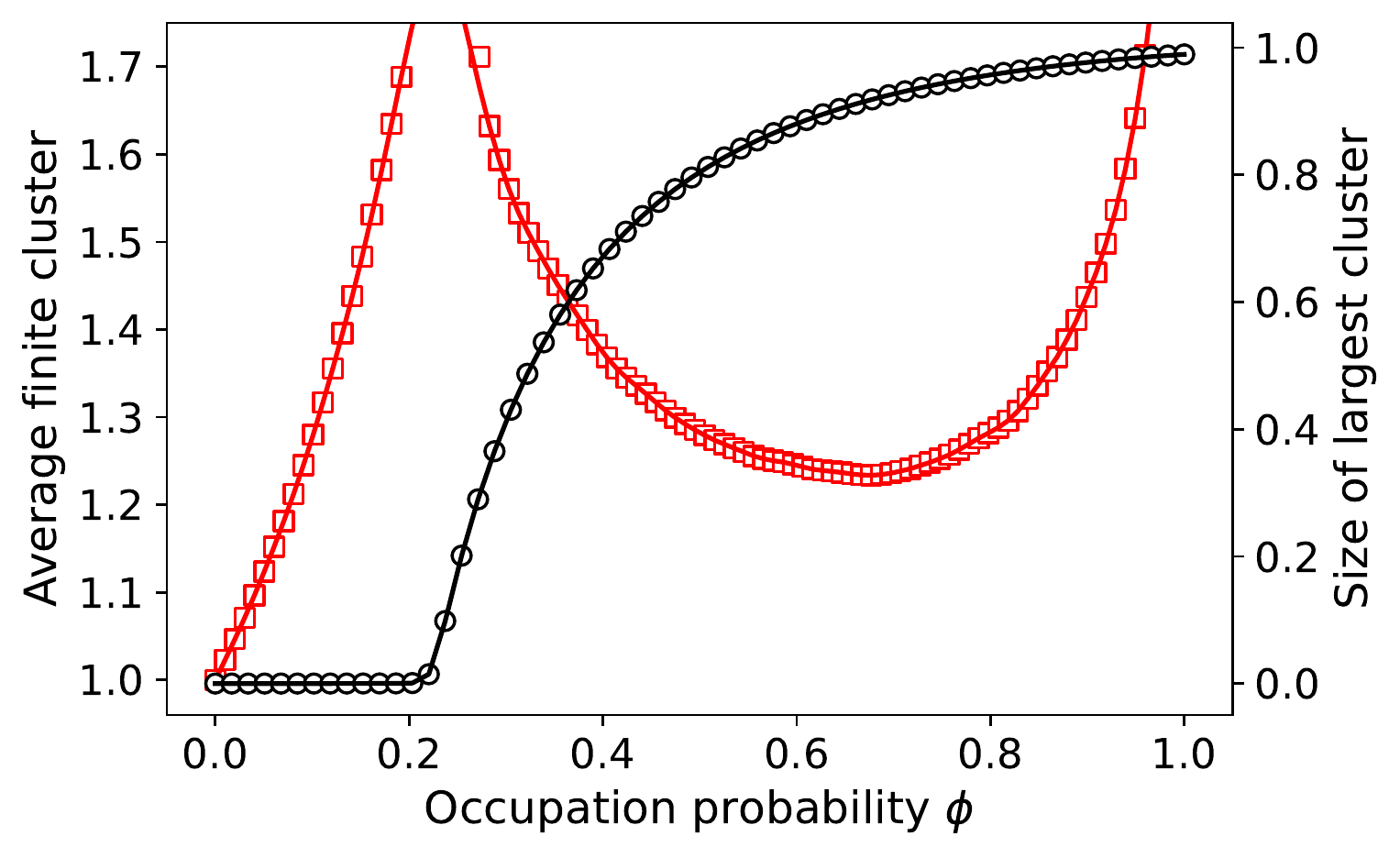}
    \caption{A random GCM graph with 2- 3- and 4-cliques. Solid lines are the results of the message passing calculations from Eq \ref{eq:S} with Eq \ref{eq:main_clique} and Eq \ref{eq:small_component} with Eq \ref{eq:main_clique_derivative}. The scatter points are the average of Monte Carlo simulations of bond percolation on the same network. Circles are the average size of the largest cluster; whilst squares are the average finite component sizes. }
    \label{fig:random_graph}
\end{figure}

The exact expressions we have derived work well for random graph models, such as the generalised configuration model \cite{karrer_newman_2010,PhysRevE.103.012309,PhysRevE.103.012313}. In Fig \ref{fig:random_graph} we apply these equations to a random graph model comprising 2-, 3- and 4-cliques that has been constructed according to the GCM algorithm; observing excellent agreement between Monte Carlo bond percolation and our theoretical model. 
In order to apply our expressions to the study of real world networks, we must find a way to cover the network in motifs. In what amounts to community detection \cite{fortunato_hric_2016}, there are perhaps a near infinite number of ways to cover a network with motifs; and some will inherently lead to better models of the empirical network than others \cite{10.5555/1036843.1036914,burgio_arenas_gomez_matamalas_2021,PhysRevE.105.044314}. The problem of network dismantling is closely related \cite{ritmeester_meyer-ortmanns_2022,li_2021,PhysRevE.103.L061302}. 

A trivial solution is to simply cover the network in 2-cliques. In doing this, we are assuming that the network is treelike; this would lead to the traditional belief propagation model, which is known to suffer from statistical errors. To progress beyond this model, the simplest loop we might add to a cover is a triangle; and beyond this chordless cycles and cliques of all orders. However, restricting the permissible motifs to cliques and cycles is unlikely to lead to a locally treelike factor graph. At the other extreme of this logic, one might try to define motifs that contain $\mathcal O(N)$ vertices; perhaps the largest Eulerian cycle for instance. Whilst this cover might lead to a treelike factor graph, we could not hope to write the message passing equations in closed form. Therefore, we must find a balance between motifs that are arbitrary enough to make the factor graph \textit{sufficiently} treelike, yet are small enough to be analytically tractable such that their message passing equations can be calculated in reasonable time.  

Let us suppose that we have defined a set of motifs, perhaps cliques of all sizes. We now have to decide a strategy of how to place the motifs on the network. By far the most simple strategy is to simply search for the presence of a pre-defined set of higher-order motifs and greedily add them to the network cover. This approach is usually stochastic if different starting locations (and subsequent search patterns) are chosen. Burgio \textit{et al} \cite{burgio_arenas_gomez_matamalas_2021} suggest that a \textit{maximal} cover is the best strategy to retain as much higher-order structure of the empirical networks as possible. In their heuristic, the cover is chosen such that the fewest cliques possible are placed on the network; minimising the local disruption caused by placing a motif. Mann \textit{et al} \cite{PhysRevE.105.044314} showed that preferentially including the \textit{largest} cliques can be beneficial for retaining the degree correlations among the more well-connected vertices; although this approach is often not maximal. Both of these methods can be generalised to placing motifs other than cliques.

\begin{figure}[ht!]
    \centering
    \includegraphics[width=0.385\textwidth]{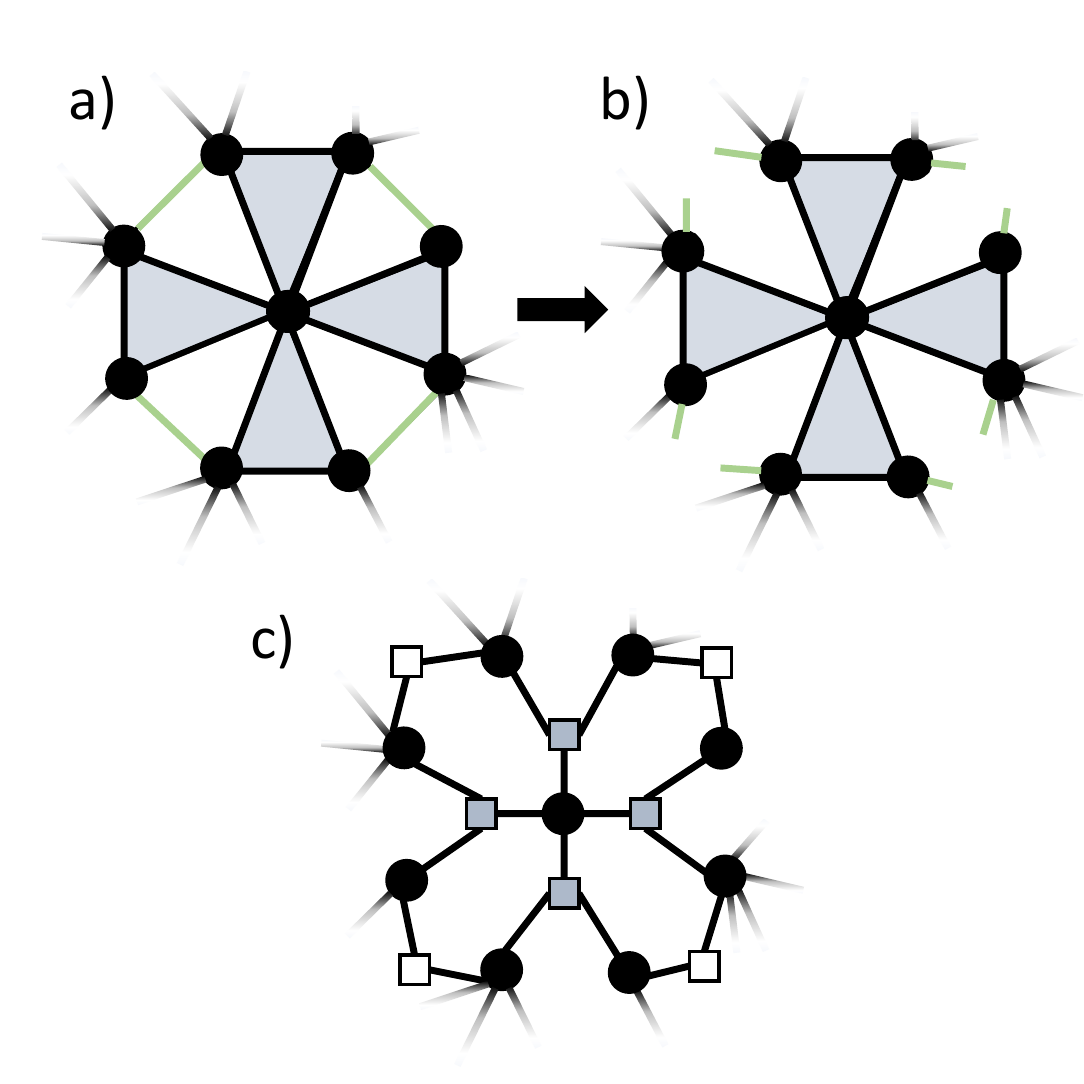}
    \caption{a) The neighbourhood of a focal vertex $i$ covered with ordinary edges (green) and triangles (shaded). Only half of the connections between the neighbours are included in the neighbourhood of $i$ due to the edge-disjoint property of the cover. b) When we calculate the cavity equations $H_{i\leftarrow \tau(z)}$ for $i$, the ordinary edges are included in the product over the motifs that vertex $ j \in\tau\backslash i$ belongs to $\prod_\nu H_{ j \leftarrow \nu}(z)$; introducing bias and therefore statistical error. c) The factor graph contains short range loops.}
    \label{fig:awkward_motif}
\end{figure}

\begin{figure}[ht!]
    \centering
    \includegraphics[width=0.475\textwidth]{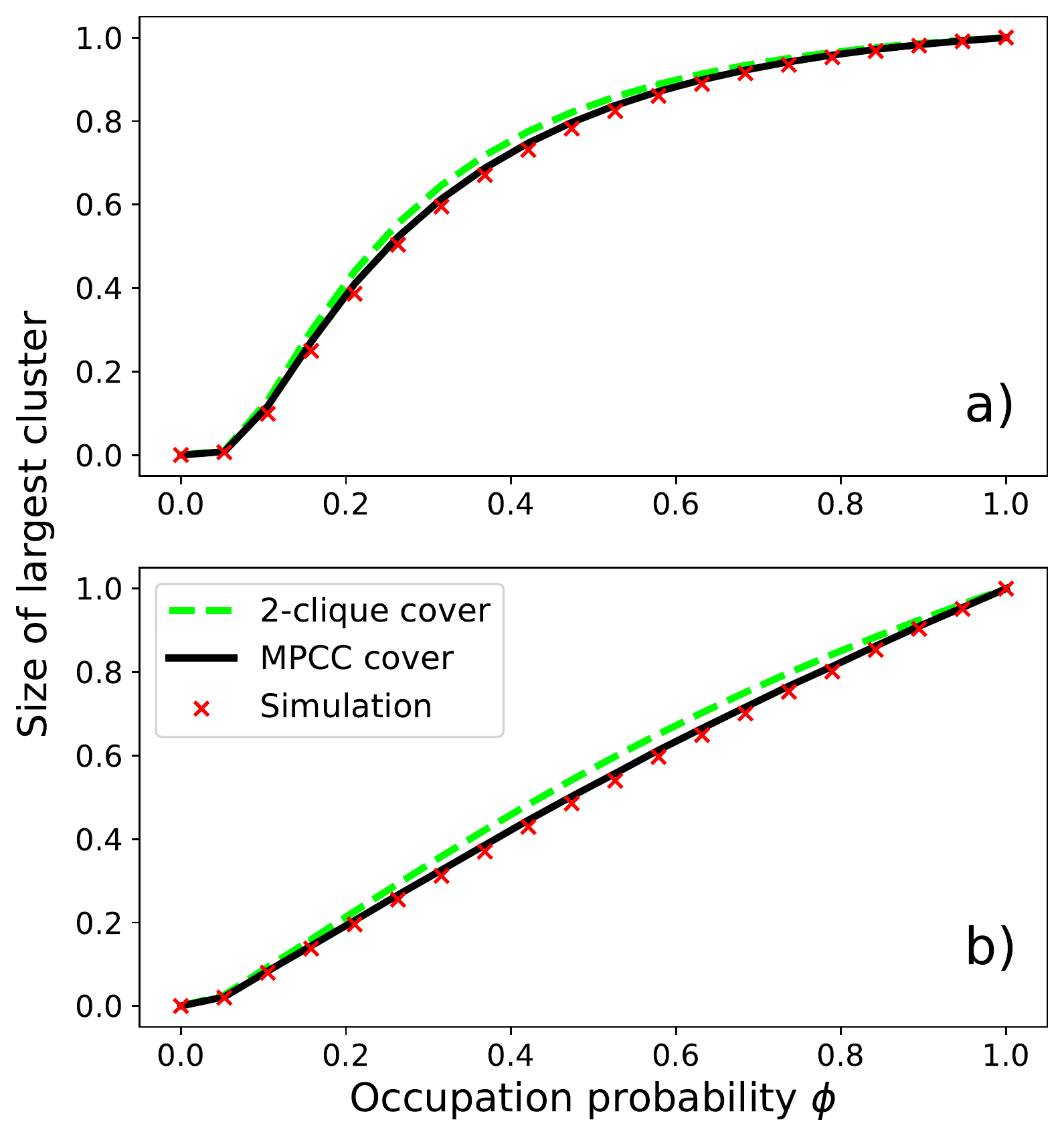}
    \caption{The size of the largest cluster of: a) a coauthorship network of 13,861 scientists \cite{doi:10.1073/pnas.98.2.404} and (b) a network of 10,680 users of the PGP encryption software \cite{PhysRevE.70.056122}. The scatter points are the average of Monte Carlo simulation of bond percolation. The dashed (green) line is the result of the message passing equations with the 2-clique cover; whilst the solid (black) line is the result for the MPCC cover and Eq \ref{eq:main_clique}.}
    \label{fig:coauthorship}
\end{figure}

Finally, we would like to highlight another concern that should be accounted for when choosing the motif cover - symmetry. Consider the neighbourhood of a vertex that has been covered by 2- and 3-cliques in Fig \ref{fig:awkward_motif} a). Due to the edge-disjoint property of the cover, only half of the edges between the neighbours are included in triangles (and therefore, motifs surrounding the focal vertex). When calculating the message passing equations, the focal vertex only observes the effective neighbourhood depicted in Fig \ref{fig:awkward_motif} b). Because of the asymmetry in how the edges between 1st-order neighbours are accounted for, this cover introduces a statistical bias due to the preferential inclusion of only \textit{some} of the edges between the 1st-order neighbours and not others. To remove the bias we must either choose the 2-clique cover or, choose the entire motif to be part of the cover. This means that clique covers, which almost always involve breaking symmetry between the neighbours, are inferior to covering methods that account for the entire neighbourhood of a vertex \cite{cantwell_newman_2019}. 
 Despite these drawbacks, our model gives good results for real world networks, see Fig \ref{fig:coauthorship}. In Fig \ref{fig:coauthorship} a), we show the size of the largest connected component for a social network of coauthorship relations between 13,861 scientists \cite{doi:10.1073/pnas.98.2.404} in the field of condensed-matter physics; and b) a network of 10,680 users of the PGP encryption software \cite{PhysRevE.70.056122}. These networks are known to contain a high-density of short loops; and therefore, the standard message passing equations, based on a 2-clique cover (dashed green), fail to correctly predict the size of the percolating cluster. To cover the networks in cliques, we use the so-called \textit{motif preserving clique cover} (MPCC) \cite{PhysRevE.105.044314}. This heuristic attempts to include the largest cliques in the edge-disjoint cover in order to preserve as many contacts among the vertices within a given motif. We find a significant improvement over the 2-clique cover, therefore highlighting the importance of retaining neighbourhood structure for the message passing theory. These networks were previously studied by Cantwell and Newman \cite{cantwell_newman_2019}. Their neighbourhood method offers closer agreement to the simulated network than our approach due to the arbitrary nature of the motifs that are included in the cover. It is, however, encouraging that simply incorporating larger cliques into the model offers an improvement over the traditional message passing theory. Cliques are particularly favourable due to the wide range of algorithms and theoretical results that have been developed for these motifs, including those of section \ref{sec:cliques_formula}.  We comment that finding the clique cover and solving the message passing equations for these networks takes just over 1 minute on 16 GB Apple M1 Silicone, which is very fast compared to other techniques. 

\section{Conclusion}
\label{sec:conclusion}

Belief propagation on loopy networks is a topic that has received much attention in the literature. In this paper, we have studied a bond percolation model on networks that contain simple cycles and cliques by deriving a message passing model. We assume that the factor graph of these networks is locally treelike in order to reduce the correlations between the messages each vertex observes. We examined our theoretical framework for random graphs constructed according to the generalised  configuration model as well as a covered empirical networks. We found excellent agreement between Monte Carlo simulation of the average cluster size distribution and the expected size of the percolating component and our equations. Our model offers significant advantage over the traditional belief propagation framework when applied to empirical networks that contain a non-trivial density of loops.

The choice of motif cover is influential to the success of the model and we highlighted some considerations around this. We conjecture that, if the set of motifs that defined the cover in our model was broad or arbitrary enough to capture the local neighbourhood of each vertex at sufficient distance, then our method would produce equivalent results to those of \cite{cantwell_newman_2019}. However, it is non-trivial to derive the conditional message probabilities for a given edge configuration of each motif in a closed form expression. Further investigation of the impact of cover choice should be conducted including: the choice of permissible motifs in the model, different cover rules for mesoscale network structures, placing large or small motifs preferentially, the hardness of the constraints in the search for the optimally treelike factor graph and constraints to find optimally symmetric neighbourhood covers.

Finally, we showed how this message passing formulation reproduces the generalised configuration model \cite{karrer_newman_2010} when it is averaged over all networks in an ensemble of graphs that have an equivalent joint degree distribution. 

These results give important insight into how the belief propagation algorithm can be applied to random and empirical networks. Further work should be carried out to find the critical point of models that contain chordless cycles and cliques \cite{PhysRevLett.113.208701,PhysRevLett.113.208702}.

\appendix

\section{Number of connected graphs $\mathcal Q_{n,k}$}
\label{sec:appendix:connected_graphs}

The number of connected graphs $\mathcal Q_{n,k}$ with $n$ vertices and $k$ edges has been discussed previously in the context of evaluating the percolation formulas for cliques \cite{PhysRevE.68.026121, PhysRevE.104.024304}. There are at least three well-known approaches to evaluating this quantity including an asymptotic expansion by Flajolet \textit{et al} \cite{flajolet_salvy_schaeffer_2004}, a closed-form expression \cite{PhysRevE.104.024304} and a fast recursive formula \cite{harary_palmer_1973} due to Harary and Palmer. The recursion is given by
\begin{equation}
    \mathcal Q_{n, k} =
    \begin{cases}
0 \qquad &k< n-1,\quad \text{or}\quad k> n(n-1)/2, \\
n^{n-2} \qquad &k = n-1, 
\quad\\
Q(n,k)\qquad &\text{otherwise}.
\end{cases}
\end{equation}
where 
\begin{align}
    Q(n,k)=&\binom{\frac 12n(n-1)}{k}
- \sum\limits_{m=0}^{n-2} {n-1\choose m} 
\sum\limits_{p=0}^k \nonumber\\
&\times{\frac 12(n-1-m)(n-2-m) \choose p} \mathcal Q_{m+1, k-p}.
\end{align}
This recursive formula can yield the coefficients of cliques with size larger than 100 vertices in fractions of a second running on 16 GB Apple M1 Silicone using Python 3.10. 

\section{ACKNOWLEDGMENTS}

This work was partially supported by the UK Engineering and Physical Sciences Research Council under grant number EP/N007565/1 (Science of Sensor Systems Software).

\subsection*{References}
\bibliography{bib}

\end{document}